\documentclass[11pt]{article}

\usepackage{graphicx}
\usepackage{latexsym}
\usepackage{bbold,bbm,dsfont}
\usepackage{setspace}
\usepackage{amssymb,amsmath}
\usepackage{mathrsfs}
\usepackage{color}

\def\thepage{\hbox to \hsize{\textcolor{blue}
{\sc Accepted at ISIT 2008 } \hfil \arabic{page}}}
\makeatletter

\newcommand{\prob}[1]{\mathsf{Pr}\left(#1\right)}

\newcommand{\Geom}{\mathrm{Geom}} 

\newcommand{\ncr}[2]{\left(\begin{array}{c}{#1} \\ #2 \end{array}
  \right)} 
\newtheorem{theorem}{Theorem}
\newtheorem{lemma}{Lemma}

\begin{document}
\title{On Distributed Function Computation in Structure-Free Random
  Networks} \author{Sudeep Kamath and D. Manjunath\footnote{Research
    carried out at the Bharti Centre for Communication Research and
    supported in part by grants from the Ministry of Information
    Technology of Government of India}\\
  Dept of Electrical Engg \\
  IIT Bombay Mumbai INDIA 400 076\\
  \texttt{sudeep,dmanju@ee.iitb.ac.in} }

\date{}
\maketitle

\begin{abstract}
  THIS PAPER IS ELIGIBLE FOR THE STUDENT BEST PAPER AWARD. We consider
  in-network computation of MAX in a structure-free random multihop
  wireless network. Nodes do not know their relative or absolute
  locations and use the Aloha MAC protocol.  For one-shot computation,
  we describe a protocol in which the MAX value becomes available at
  the origin in $O(\sqrt{n/\log n})$ slots with high probability. This
  is within a constant factor of that required by the best coordinated
  protocol. A minimal structure (knowledge of hop-distance from the
  sink) is imposed on the network and with this structure, we describe
  a protocol for pipelined computation of MAX that achieves a rate of
  $\Omega(1/(\log^2 n)).$
\end{abstract}

\section{Introduction}
\label{sec:intro}
Early work on computation of functions of binary data over wireless
networks focused on computing over noisy, time-slotted, broadcast
networks, e.g., \cite{Gallager88,Kushilevitz98}. With increasing
interest in wireless sensor networks, recent research has concentrated
on `in-network' computation over multihop wireless networks, e.g.,
\cite{Giridhar05,Khude05,Kanoria07}, which also assume that time is
slotted.  The primary focus of the above research has been to define
an oblivious protocol that identifies the nodes that are to transmit
in each slot.  This implies that the nodes have organized themselves
into a network and have their clocks synchronized. Both of these
require significant effort. In this paper we describe a protocol for
in-network computation of \texttt{MAX} in a structure-free network,
(i.e., in a network where nodes do not have an identity and hence do
not know the topology) that uses the Aloha MAC protocol. We first
describe the \textsf{One-Shot MAX} protocol for one-shot computation
of the \texttt{MAX} and its analysis. We show that, with high
probability (w.h.p.), the sink will have the result in a time that is
within a constant factor of that required by a structured network.  We
then impose a minimal structure and describe the \textsf{Pipelined
  MAX} protocol and its analysis. We show that the rate of computing
the \texttt{MAX} in this network is $\Omega(\frac{1}{\log^2 n}).$

\section{\texttt{MAX}  in Multihop Aloha }
\label{sec:MAX in Multihop Aloha}
$n$ nodes are uniformly distributed in $[0,1]^2$ and each node is
assumed to know $n.$ The sink, the node that is to have the value of
the \texttt{MAX}, is at the origin. The nodes do not have an identity
and they do not know either their relative or their absolute
positions. Hence, the network does not know its topology. This of
course means that a schedule for transmissions cannot be defined. Thus
a random access protocol is an obvious choice at the MAC layer. We
first assume that the nodes use the s-Aloha MAC protocol. For
pedagogical convenience, we will assume slotted-Aloha at the MAC
layer.  The analysis easily extends to the case of pure Aloha MAC.

Spatial reuse is analyzed using the well-known protocol model of
interference \cite{Gupta98}. For s-Aloha, this model translates to the
following. Consider a transmitter at location $x_1$ transmitting in a
slot $t.$ A receiver at location $x_2,$ can successfully decode this
transmission if and only if the following two conditions are
satisfied. (1) $\| x_2 - x_1 \| < r_n,$ and (2) $\| x_2 - x_3 \| >
(1+\Delta^\prime)r_n$ for some constant $\Delta^\prime \geq 0;$ $x_3$
is the location of any other node transmitting in slot $t.$ $r_n$ is
called the transmission radius. A transmission in slot $t$ is deemed
successful if all nodes within $r_n$ of the transmitter receive it
without collision. The following is a sufficient condition for
successful transmission by a node located at $x$ in a slot: $\|x -
x^\prime \| > (1 + \Delta)r_n,$ $\Delta = 1 + \Delta^\prime,$ for
every other node transmitting in that slot and located at $x^\prime.$

\subsection{One-shot computation of \texttt{MAX} using Aloha}
\label{sec:one-shot}
Let $Z_i$ be the value of the one-bit data at Node~$i$ and
$\mathcal{Z} := \max_{1 \leq i \leq n} Z_i.$ The protocol
\textsf{One-Shot MAX} is as follows. Node $i$ can either receive or
transmit in a slot but not both. In slot $t,$ Node~$i$ will either
transmit, with probability $p$ or listen, with probability $(1-p),$
independently of all the other transmissions in the network.  Let
$X_i(t)$ be the value of the bit received (i.e., correctly decoded in
the absence of a collision) by Node~$i$ in slot $t,$ $t=1,2,\ldots.$
If Node~$i$ transmits in slot $t$ or if it senses a collision or idle
in the slot, then it sets $X_i(t) = 0.$ Define $Y_i(0) = Z_i$ and
$Y_i(t):=\max \{ Y_i(t-1), X_i(t)\}$ for $t=1,2, \ldots.$ $Y_i(t)$ is
the `running MAX' at Node~$i$ in slot $t.$ If Node~$i$ transmits in
slot $t,$ it will transmit $T_i(t)=Y_i(t-1).$

It is easy to see that the correct value of $\mathcal{Z}$ will
`diffuse' in the network in every slot. The performance of the
protocol, that is, the diffusion time, depends on $p.$ The choice of
$p$ is discussed in Section~\ref{sec:proofs}.

To study the progress of the diffusion, we will consider a
tessellation of the unit square into square cells of side $s_n =
\lceil \sqrt{\frac{n}{2.75\log n}}\rceil^{-1}.$ This will result in
$l_n :=\frac{1}{s_n} = \lceil\sqrt{\frac{n}{2.75\log n}}\rceil$ rows
(and columns) of cells in $[0,1]^2.$ There will be a total of $M_n :=
\frac{1}{s_n^2}=\lceil\sqrt{\frac{n}{2.75\log n}}\rceil^{2}$ cells.
Let $\mathcal{C}$ denote the set of cells under this tessellation. Let
$S_c$ be the set of nodes in Cell~$c$ and $N_c$ be the number of nodes
in Cell~$c.$ Under this tessellation, two cells are said to be
\textit{adjacent} if they have a common edge. Let the transmission
radius be $r_n = \sqrt{\frac{13.75 \log n}{n}} \approx \sqrt{5}s_n.$
For this value of $r_n$ the network is connected w.h.p.
\cite{Gupta98}.  The expected number of nodes in a cell is $n s_n^2
\approx 2.75\log n.$ Further, from Lemma~3.1 of \cite{Xue04}, for our
choice of $r_n$ and $s_n,$
\begin{equation}
  \label{eq:bounds on N_c}
  \prob{c_1 \log n \leq N_c \leq c_2 \log n\ \mbox{for}\ 1 \leq c \leq 
      M_n}  \ \to \ 1
\end{equation}
where $c_1 = 0.091$ and $c_2 = 5.41.$ Our results will hold for
networks that are connected and which satisfy \eqref{eq:bounds on
  N_c}. From the choice of $r_n$ (i.e., $r_n\geq\sqrt{5}s_n$) a
successful transmission by any node from Cell~$c$ is correctly decoded
by all nodes in Cell~$c$ as well as by all nodes in cells adjacent to
Cell~$c.$

\begin{figure}
  \begin{center}
    \includegraphics[width=2.7in]{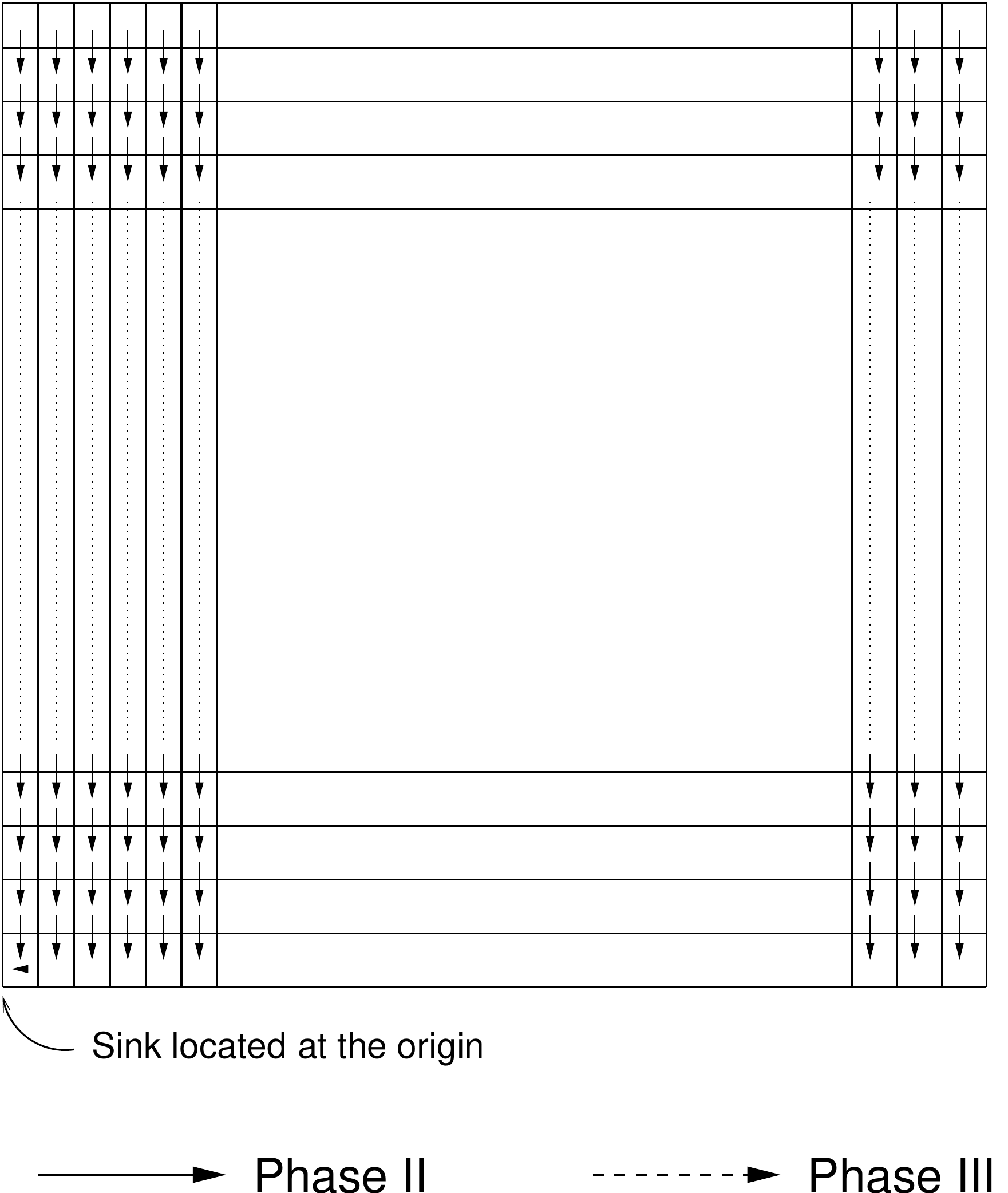}
    \caption{Direction of diffusion during Phase II and Phase III of 
      protocol \textsf{One-Shot MAX}}
    \label{fig:square}
  \end{center}
\end{figure}

The value of $\mathcal{Z}$ can reach the sink along any of the many
possible trees rooted at the sink. For our analysis, we will divide
the progress of the diffusion into the following three `phases' and
analyze each of the three phases separately. We reiterate that the
above sequence of phases is for the purpose of analysis of the time to
diffusion. The nodes do not perform any such organization.

\begin{itemize}
\item \textit{Phase I for data aggregation within each cell.} This
  phase is completed when every node of the network has transmitted
  successfully at least once.
\item \textit{Phase II for progress to the bottom of the square.} In
  this phase, the locally computed values of the \texttt{MAX} get
  diffused into the cells on one side of the unit square as shown in
  Fig. ~\ref{fig:square}.
\item \textit{Phase III for progress into the sink.} In this phase,
  the value of \texttt{MAX} reaches the sink at the origin in the
  manner shown in Fig.~\ref{fig:square}.
\end{itemize}

We show in Section~\ref{sec:proofs} that Phase I will be completed in
$O(\log^2 n)$ slots w.h.p., Phase II and Phase III will
each be completed in $O\left(\sqrt{\frac{n}{\log n}}\right)$ slots
w.h.p. These results are combined into the following
theorem.

\begin{theorem}
  If all the nodes execute the protocol \textsf{One-Shot MAX}, then
  for any $\alpha, k > 0,$ the maximum of the binary data at the $n$
  nodes is available at the sink with probability at least
  $\left(1-\frac{k}{n^\alpha}\right)$ in $O\left(\sqrt{\frac{n}{\log
        n}}\right)$ slots.
  \label{thm:one-shot}
\end{theorem}

We will also argue that in another $O\left(\sqrt{\frac{n}{\log
      n}}\right)$ number of slots, the value of $\mathcal{Z}$ would
have diffused to each node of the network. Note that the best one-shot
protocol in an organized network, under this choice of $r_n$ will also
require $\Theta\left(\sqrt{\frac{n}{\log n}}\right)$ time slots for a
one-shot computation of \texttt{MAX}. The bound on the time in
Theorem~\ref{thm:one-shot} is therefore tight.

\subsection{Pipelined computation of \texttt{MAX} using Aloha}
\label{sec:pipelined}

If $\mathcal{Z}$ were to be computed continuously using the
\textsf{One-Shot MAX} protocol, a throughput of
$\Theta\left(\sqrt{\frac{\log n}{n}}\right)$ can be achieved. We
believe some structure in the network is necessary to do better.  We
will assume that all nodes have a transmission range that is exactly
$r_n.$ This strict requirement can be easily relaxed but we will keep
this assumption for pedagogical convenience.

We impose the following structure in the network. Prior to the
computation, each node obtains its minimum hop distance to the sink.
Henceforth, we will refer to this as simply the \textit{hop distance}
of the node. From \eqref{eq:bounds on N_c}, each cell in the
tessellation is occupied.  Since nodes in adjacent cells differ in
their hop distance by atmost $1,$ the largest hop distance of a node
in the network is no more than $d:= 2\,l_n = 2\lceil
\sqrt{\frac{n}{2.75\log n}}\rceil.$

Let $h_i$ be the hop distance of Node~$i.$ Observe that a transmission
by Node~$i$ can be decoded successfully by Node~$j$ only if
$|h_i-h_j|\leq 1.$ Hence, if there is a reception by Node~$i$ in slot
$t,$ then that transmission must have been made by a node with hop
distance either $(h_i-1),\ h_i,$ or $(h_i+1).$ Thus, if a node
transmits its hop distance modulo $3$ along with its transmitted bit,
then every receiver that can decode this transmission successfully,
can also, by the receiver's knowledge of its own hop distance,
correctly identify the hop distance of the transmitter.

Time is divided into \textit{rounds}, where each round consists of
$\tau$ slots. Minimizing $\tau$ would maximize the throughput. We will
discuss this in Section~\ref{sec:proofs}. Data arrives at each node at
the beginning of each round, that is, at the rate of $1$ data bit per
round. Let the value of the bit at Node~$i$ in the round $r$ be
$Z_i(r).$ $\mathcal{Z}(r) := \max_{1 \leq i \leq n} Z_i(r),$ for
$r=1,2,\ldots,$ is to be made available at the sink node, Node~$s.$

\textsf{Pipelined MAX} protocol is as follows. The sink only receives
data and does not transmit. The other nodes in the network perform the
following. (We remind here that the naming of the nodes is for our
convenience. The nodes themselves do not know their identity.)

In each slot, Node~$i$ either transmits with probability $p$ or
listens with probability $(1-p)$ independently of all other
transmissions in the network. The value of $p$ is chosen as in the
\textsf{One-Shot MAX} protocol. Each node executes the following
protocol for round $r.$

\textit{Transmission:} If Node~$i$ transmits in slot $t$ of round $r$,
then it transmits three bits $(A_i, B_i, T_i(r,t))$ in the slot. Bits
$A_i$ and $B_i$ are the \textit{identification bits} and are obtained
as $(h_i \bmod 3).$ The bit $T_i(r,t)$ is the transmitted \textit{data
  bit} and is obtained as
\begin{displaymath}
  T_i(r,t)=\max\{Z_i(r-d+h_i), Y_i(r-1)\}. 
\end{displaymath}
Here, by convention, $Z_i(v)=Y_i(v)=0$ for $v\leq 0.$ $Y_i(r-1)$ is
computed from succesful receptions in round $(r-1),$ as described
below.

\textit{Reception:} In round $r,$ Node~$i$ maintains $Y_i(r,t)$ for
$t=0,1,2,\ldots,\tau.$ $Y_i(r,0)$ is initialized to $0$ at the
beginning of round $r.$ $Y_i(r,t)$ stores the \texttt{MAX} of the data
bits that Node~$i$ has decoded from all the slots in round $r,$ upto
and including slot $t,$ and which were transmitted by the nodes with
hop distance $(h_i+1).$ In slot $t$ of round $r,$ if Node~$i$
successfully receives a transmission from a node with hop distance
$(h_i+1)$ (available from the identification bits), then it sets the
received data bit to be $X_i(r,t).$ If Node~$i$ senses an idle or a
collision in slot $t,$ or if it receives a successful transmission
from a node with hop distance different from $(h_i+1),$ then it sets
$X_i(r,t) = 0.$ Thus, $Y_i(r,t) = \max \{ Y_i(r,t-1), X_i(r,t) \}.$
Define $Y_i(r):=Y_i(r,\tau).$

The sink node, Node~$s,$ obtains the \texttt{MAX} as
$\mathcal{Z}(r-d)=\max\{Z_s(r-d),Y_s(r)\},$ for all $r > d.$ The delay
of the protocol is $d$ rounds or $d \tau $ time slots.

\begin{theorem}

  If all the nodes execute the protocol \textsf{Pipelined MAX}, then
  for any $\alpha, k > 0,$ there exists $\tau = \tau(\alpha,k) =
  \Theta(\log^2 n)$ so that the correct \texttt{MAX} is available at the sink
  in a round with probability atleast
  $\left(1-\frac{k}{n^\alpha}\right).$ This achieves a throughput of
  $\Omega\left(\frac{1}{\log^2 n}\right)$ with a delay of
  $O(\sqrt{n\log^3 n})$ slots.
  \label{thm:pipelined}
\end{theorem}

The optimal pipelined protocol for \texttt{MAX} in an organized
network requires $\Theta(\log n)$ slots for each round in the absence
of block coding. Thus, the penalty for minimal organization and no
coordination is the $\log n$ overhead for the length of each round.
Also, for our protocol, Node~$i,$ with a hop distance of $h_i,$
requires a memory of $(d-h_i+1)$ bits to store $Z_i(r), Z_i(r-1),
\ldots, Z_i(r-d+h_i).$ Thus, the protocol requires each node to have
$(d+1)$ bits of memory for storage of past data values.

\section{Proofs}
\label{sec:proofs}

\subsection{Preliminaries}

\subsubsection{Bounding the Number of Interfering Neighbors}

Define the interfering neighborhood of Node~$i$ by
${\mathcal{N}}^{(I)}_i := \{ j: 0 < \| X_i - X_j\| \leq (1+\Delta)r_n
\}.$ As discussed earlier, a transmission from Node~$i$ in slot $t$ is
deemed successful if all nodes within $r_n$ of Node~$i$ can decode
this transmission without a collision. A sufficient condition for
Node~$i$ to be successful in transmitting in slot $t$ is that no node
belonging to ${\mathcal{N}}^{(I)}_i$ must transmit in slot $t.$

From the protocol model, the choice of $s_n$ and \eqref{eq:bounds on
  N_c}, the set of nodes that interfere with a transmission from a
node in Cell~$c$, (i.e., $\bigcup_{i\in S_c}{\mathcal{N}}^{(I)}_i$) is
contained within an interference square centered at Cell~$c.$ This
square contains $k_1 = \left(2\lceil\frac{(1+\Delta)r_n}{s_n}\rceil +
  1\right)^2$ cells. From \eqref{eq:bounds on N_c},
\begin{equation}
  \label{eq:NIi bound}
  |{\mathcal{N}}^{(I)}_i| \leq k_1 c_2 \log n - 1
\end{equation} 
Observe that $k_1$ is a constant for large enough $n.$

\subsubsection{Probability of a successful transmission from a cell}
Let $P_i$ be the probability that Node~$i$ transmits successfully in a
slot and $P^{(c)},$ the probability that some node in Cell~$c$
transmits successfully in a slot.  $P_i \geq
p(1-p)^{|{\mathcal{N}}^{(I)}_i|},$ and from \eqref{eq:NIi bound}, we
have $P_i \geq p(1-p)^{k_1c_2\log n-1}.$ Successful transmissions by
nodes from Cell~$c$ are mutually disjoint events, and hence, $P^{(c)}
= \sum_{i\in S_c} P_i \geq N_cp(1-p)^{k_1c_2\log n-1}.$ From
\eqref{eq:bounds on N_c}, we have $N_c \geq c_1\log n\ \forall c\in
\mathcal{C}$ and hence, $P^{(c)} \geq c_1\log n\ p(1-p)^{k_1c_2\log
  n-1}.$ Choosing $p = \frac{1}{k_1c_2\log n}$ maximises the lower
bound in this inequality and yields

{\footnotesize
\begin{displaymath}
  P^{(c)} \ \geq \ \frac{c_1}{k_1c_2} \left(1+\frac{1}{k_1c_2\log
      n-1}\right)^{-(k_1c_2\log n-1)} \ \geq \ \frac{c_1}{k_1c_2e} =:
  p_S 
\end{displaymath}
}

\normalsize
Thus, the probability of successful transmission from a cell is lower
bounded by a constant $p_S,$ independent of the number of nodes in the
network. This will be crucial to our analysis.

\subsection{Proof of Theorem~\ref{thm:one-shot}}
We will prove Theorem~\ref{thm:one-shot} by proving bounds on the
total time required by each of phases I, II and III.

\subsubsection{Phase I: Data aggregation within each cell}
Consider Cell~$c.$ Let $\mathcal{T}_c$ be the total number of slots
required for every node in Cell~$c$ to have transmitted successfully
atleast once. Recall that $p= (k_1c_2\log n)^{-1}.$ We will bound
$\mathcal{T}_c$ by stochastic domination. Consider a sample space
$\mathcal{S}$ containing mutually disjoint events $E_1, E_2, \ldots,
E_{N_c}.$ Let $\prob{E_q}=p(1-p)^{k_1c_2\log n-1}$ for $1\leq q \leq
N_c.$ Observe that $P_i \geq \prob{E_q}\ \forall i\in S_c$ and $1\leq
q\leq N_c.$ Let $E=\bigcup_{q=1}^{N_c} E_q.$ We have $P_E := \prob{E}
= N_cp(1-p)^{k_1c_2\log n-1}.$ Let a sequence of samples be drawn
independently from $\mathcal{S}.$ Let the number of samples required
to be drawn from $\mathcal{S}$ so that each of the events $E_q,\
q=1,2,\ldots N_c$ occurs atleast once, be the random variable
$T^\prime_c.$ The probability of occurence of $E$ in a given sample is
$P_E$ and hence, the waiting time in terms of number of samples drawn,
for the event $E$ to occur, as well as the waiting time between
consecutive occurences of $E,$ is given by the geometrically
distributed random variable $\Geom(P_E).$ Now, consider the events of
successful occurences of event $E.$ If $(l-1)$ distinct events among
$E_q, 1\leq q\leq N_c$ have already occured, then the probability that
the next occurence of $E$ is due to an as yet unoccured event
$E_{q^\prime}$ is $(1 - \frac{l-1}{N_c}),$ as each $E_q, 1\leq q\leq
N_c$ is equally probable. The number of occurences of event $E$ to
wait for the occurence of an as yet unoccured event among $E_q, 1\leq
q\leq N_c$ is distributed as $\Geom(1 - \frac{l-1}{N_c}).$ The random
variable $T^\prime_c$ can thus, be expressed as: $T^\prime_c =
\sum_{j=1}^{R^\prime_c} t^\prime_{c,j}$ where $t^\prime_{c,j} \sim
\Geom(P_E)$ and $R^\prime_c \sim \sum_{l=1}^{N_c} \Geom(1 -
\frac{l-1}{N_c}).$

Now compare the following two events: (1) Event $\mathcal{A}$ defined
as the successful transmission from Cell~$c$ resulting from a
successful transmission by Node~$i$ in Cell~$c$ and (2) Event
$\mathcal{B}$ defined as the occurrence of $E$ in a sample drawn from
$\mathcal{S}$ due to the occurence of $E_q.$ Observe that
$\prob{\mathcal{A}} \geq \prob{\mathcal{B}}.$ From this comparison, we
see that $\mathcal{T}_c$ will be stochastically dominated by
$T^\prime_c$ i.e. $\prob{\mathcal{T}_c\geq z}\leq \prob{T^\prime_c\geq
  z}\ \forall z\in\mathbb{N}$. Further, $T^\prime_c$ will be
stochastically dominated by the random variable $T_c =
\sum_{j=1}^{R_c} t_{c,j},$ where $t_{c,j} \sim \Geom(p_S)$ and $R_c
\sim \sum_{l=1}^{m} \Geom(1-\frac{l-1}{m})$ with $m = \lceil c_2\log
n\rceil$ which is an upper bound on $N_c$ from \eqref{eq:bounds on
  N_c}. We therefore, have
\begin{displaymath}
  \prob{\mathcal{T}_c\geq z}\leq \prob{T_c\geq z}\ \forall z\in\mathbb{N}
\end{displaymath}
It is convenient to work with the random variable $T_c$ because it is
independent of the parameters of Cell~$c.$ We will obtain the moment
generating functions (mgf) of the distributions of the integer-valued
random variables involved.  Let the mgf of each random variable be
denoted by the same character in sans serif font. For a random
variable $F,$ $\mathsf{F}(z) = \sum_{j\in\mathbb{Z}}\prob{F=j}z^{-j}.$
The region of convergence of the mgf is specified in parentheses.
\begin{eqnarray*}
  \mathsf{t_{c,j}}(z) & = & \frac{p_S z^{-1}}{1-(1-p_S)z^{-1}}:=S(z)\
  \ \left(|z|>1-p_S\right) \\
  \mathsf{R_c}(z) & = & \Pi_{l=1}^m \frac{(1 -
    \frac{l-1}{m})z^{-1}}{1 -\frac{l-1}{m}z^{-1}}
   \ \ \left(|z|>1-\frac{1}{m}\right) \\
  \mathsf{T_c}(z) & = & \sum_{r\in\mathbb{N}} \prob{R_c=r}[S(z)]^r \\
  & = & \mathsf{R_c}\left(\frac{1}{S(z)}\right) \\
  & = & \frac{m!p_S^m}{\Pi_{l=1}^m\left(m[z-(1-p_S)] -
      (l-1)p_S\right)} \\ 
  &   &\hspace{100pt} \left(|z| > 1 - \frac{p_S}{m}\right)
\end{eqnarray*}
Thus, $\mathbb{E}[e^{sT_c}] =
\frac{m!p_S^m}{\Pi_{l=1}^m\left(m[e^{-s}-(1-p_S)] - (l-1)p_S\right)}$
for $s<\log\left(\frac{1}{1-\frac{p_S}{m}}\right).$ Choose
$s_1=\log\left(\frac{1}{1-\frac{p_S}{2m}}\right).$ After some algebra,
we can show the following.
\begin{eqnarray*}
  \mathbb{E}[e^{s_1T_c}] & = & \frac{m!p_S^m}{m^m} \Pi_{l=1}^m
  \left(e^{-s_1}-1+\frac{m-l+1}{m}p_S\right)^{-1} \\
  & = & c_m\sqrt{\pi m}. 
\end{eqnarray*}
Here $c_m = \frac{2^{2m}}{\ncr{2m}{m}\sqrt{\pi m}} \to 1$ as
$m\to\infty$ by the Stirling approximation. From the Chernoff bound we
get $\prob{\mathcal{T}_c\geq V_1} \leq \prob{T_c\geq V_1} \leq
c_m\sqrt{\pi m}\left(1-\frac{p_S}{2m}\right)^{V_1}.$ By the union
bound, we have
\begin{displaymath}
  \prob{\max_{c\in\mathcal{C}}\mathcal{T}_c\geq V_1} \leq M_nc_m\sqrt{\pi
    m}\left(1-\frac{p_S}{2m}\right)^{V_1}
\end{displaymath}
To achieve $\prob{\max_{c\in\mathcal{C}}\mathcal{T}_{c}\geq V_1}\leq
\frac{k}{n^\alpha}$, it is sufficient to have $(1-\frac{p_S}{2m})^{V_1}
\leq \frac{k}{n^\alpha M_nc_m\sqrt{\pi m}}$ or
{\footnotesize
\begin{displaymath}
  V_1 \geq \frac{\frac{1}{2}\log m + \log M_n + \alpha\log n
    - \log k + \frac{1}{2}\log \pi + \log c_m}{-\log(1-\frac{p_S}{2m})}
\end{displaymath} 
}\normalsize Here, $m=\lceil c_2\log n\rceil,\ M_n =
\lceil\sqrt{\frac{n}{2.75\log n}}\rceil^{2}.$ Writing
$-\log\left(1-\frac{p_S}{2m}\right) = \frac{p_S}{2m} +
\frac{p_S^2}{2(2m)^2} + \ldots,$ we can see that there exists a choice
of $V_1=O(\log^2 n),$ which would be sufficient for the completion of
Phase I, i.e., every node in every cell of the network would have
successfully transmitted at least once in $V_1$ slots, with
probability at least $\left(1 - \frac{k}{n^\alpha}\right).$

\subsubsection{Phase II: Progress to the bottom of the square}
Let the columns of cells shown in Fig.~\ref{fig:square} be numbered
$C_1, C_2, \ldots, C_{l_n}.$ Let the $l_n$ cells in each column be
numbered from $1$ to $l_n$ from top to bottom. In this phase, we are
concerned with transmissions in the top $w:=l_n-1$ cells of each
column. In Phase I, each node has successfully received the
transmissions by every other node in its cell. Hence, Phase II will be
completed if the following sequence of events occurs for each column
$C$: A successful transmission by some node in the first cell of the
column, followed by a successful transmission by some node in the
second cell of the column and so on until a successful transmission by
some node in the $w$-th cell of the column.

Let the number of slots required for this sequence of events be
${\mathcal{T}}^{(C)}$ for column $C.$ We can see that
${\mathcal{T}}^{(C)}$ will be stochastically dominated by $T^{(C)} :=
\sum_{j=1}^{w} t_j^{(C)},$ where $t_j^{(C)}\sim\Geom(p_S).$ We can
thus derive the following.
\begin{eqnarray*}
  \mathsf{T^{(C)}}(z) &=&
  \frac{p_S^{w}z^{-w}}{(1-(1-p_S)z^{-1})^{w}} \\
  & &  \hspace{30pt} \left(|z|>1-p_S \right)  \\
  \mathbb{E}[e^{sT^{(C)}}] &=& \frac{p_S^w}{(e^{-s}-(1-p_S))^w} \\
  && \hspace{30pt} \mbox{for $s<\log\left(\frac{1}{1-p_S}\right)$} \\
  \prob{T^{(C)} \geq V_2} &\leq& 
  \frac{\mathbb{E}[e^{s_2T^{(C)}}]}{e^{s_2V_2}} = 2^w(1-\frac{p_S}{2})^{V_2}\\
  \prob{\max_{1\leq j\leq l_n}\mathcal{T}^{(C_j)} \geq V_2} & \leq & 
  l_n2^w(1-\frac{p_S}{2})^{V_2}
\end{eqnarray*}
where we have used $s_2 = \log(\frac{1}{1-\frac{p_S}{2}})$ in the
Chernoff bound. Thus, to achieve $\prob{\max_{1\leq j\leq
    l_n}T^{(C_j)}\geq V_2} \leq \frac{k}{n^\alpha},$ it suffices to
have $(1-\frac{p_S}{2})^{V_2} \leq \frac{k}{n^{\alpha}l_n2^w}$ or

$$V_2 \geq \frac{\alpha\log n + \log l_n + w\log 2 - \log
  k}{-\log(1-\frac{p_S}{2})}$$

Now, $l_n = \lceil\sqrt{\frac{n}{2.75\log n}}\rceil=w+1,$ and hence, $V_2 =
O\left(\sqrt{\frac{n}{\log n}}\right)$ slots are sufficient for the
completion of Phase II with probability at least
$\left(1-\frac{k}{n^\alpha}\right).$

\subsubsection{Phase III: Progress into the sink}
Phase III comprises diffusion of the \texttt{MAX} into the cell
containing the sink. Let the time required for this to happen be the
random variable $T_s.$ It is easily seen from the analysis of the
sequence of transmission for Phase II that $\prob{T_s \geq V_3}\leq
2^w(1-\frac{p_S}{2})^{V_3}$ where $w$ is as defined before.
Calculations similar to those in the analysis for Phase II show that
$V_3 = O\left(\sqrt{\frac{n}{\log n}}\right)$ slots are sufficient for
completion of this phase with probability at least
$\left(1-\frac{k}{n^\alpha}\right).$

\subsubsection{Bound on the overall time}

Since each of phases I, II and III get completed in
$O\left(\sqrt{\frac{n}{\log n}}\right)$ time slots with probability at
least $\left(1-\frac{k^\prime}{n^\alpha}\right),$ for appropriate
constants $k^\prime,$ the protocol \textsf{One-Shot MAX} achieves
computation of the \texttt{MAX} at the sink in
$O\left(\sqrt{\frac{n}{\log n}}\right)$ number of time slots with
probability at least $\left(1-\frac{k}{n^\alpha}\right).$ If the
protocol is followed for another $V_3+V_2$ slots, the true \texttt{MAX}
will diffuse to the complete bottom row, and then to the complete
network, the direction of diffusion being opposite to that in Phase
III and Phase II respectively.

\subsection{Obtaining the Hop Distance}
The following algorithm \textsf{Hop Distance Compute} obtains the hop
distance for each node in the network. $\lceil \log d \rceil$ slots
are grouped into a \textit{frame} and $\tau= \Theta(\log^2 n)$ ($\tau
= V_1$ as obtained in Phase~I analysis of protocol \textsf{One-Shot
  MAX}) frames form a \textit{superframe.}  The algorithm ends after
$(d+1)$ superframes.

Let the superframes be denoted by $g_0,g_1,\ldots,g_d.$ A node either
transmits in every slot of a frame or it does not transmit in any slot
of the frame. Each transmission is a number expressed in $\lceil\log
d\rceil$ bits.  At the beginning of the algorithm, the sink transmits
the number $0$ expressed in $\lceil\log d\rceil$ bits in each frame of
superframe $g_0.$ Each node of the network other than the sink
executes the following algorithm.  Node~$i$ makes no transmission till
it has decoded a transmission successfully. Let the first successful
reception by Node~$i$ happen in a frame belonging to superframe $g_i$
and let the decoded transmission correspond to the number $n_i$
expressed in $\lceil\log d\rceil$ bits. Node~$i$ sets its hop distance
to $(n_i+1)$ and ignores other successfully received bits in frames
from superframe $g_i.$ During the $\tau$ frames from superframe
$g_{i+1},$ Node~$i$ transmits, in each frame, the number $(n_i+1)$
expressed in $\lceil\log d\rceil$ bits, with probability $p,$
independently of all the other transmissions in the network and makes
no transmission with probability $(1-p).$ After the end of round
$g_{i+1},$ Node~$i$ makes no more transmissions. The total number of
slots required is $(d+1) \tau \lceil\log d\rceil.$

\begin{lemma}
  The nodes of the network correctly compute their minimum hop
  distance from the sink, using \textsf{Hop Distance Compute} in
  $O(\sqrt{n\log^5 n})$ time slots with probability at least
  $\left(1-\frac{k}{n^\alpha}\right)$ for any positive $\alpha$ and
  some constant $k.$
  \label{lemma:hop-distance-compute}
\end{lemma}

We omit the proof of this lemma.

\subsection{Proof of Theorem~\ref{thm:pipelined}}

Let the set of nodes at hop distance $h$ be $G_h.$ Let $t_{i,r}$ be
the first slot in round $r$ that Node~$i$ transmits succesfully in.
The number of slots in a round is $\tau= \Theta(\log^2 n)$ ($\tau =
V_1$ from Phase~I). Every node in the network would have transmitted
successfully at least once in each round of $\tau$ slots w.h.p. Let
$h_{\mbox{max}}\leq d$ be the largest hop distance of a node in the
network. In the proof, we will assume that each node of the network
transmits successfully in each round at least once. We claim that
\begin{displaymath}
\max_{i\in G_h}T_i(r,t_{i,r}) = \max_{j\in\bigcup_{h\leq
    f\leq d}G_f}Z_j(r-d+h)
\end{displaymath} 
for $0\leq h\leq h_{\mbox{max}}$ and $r> d-h.$ The sink being at hop
distance $0,$ proving the claim will complete the
proof. 
Assume that the claim is true for $h_0<h\leq h_{\mbox{max}}$ for
$r>d-h.$ We shall show that the claim will then be true for $h=h_0$
and for $r>d-h_0.$ Consider transmissions by the nodes at hop distance
$h_0$ in round $(r+1).$

{\footnotesize
 \begin{eqnarray*}
   &&\hspace{-20pt}   \max_{i\in G_{h_0}}T_i(r+1,t_{i,r+1}) =  \max_{i\in
     G_{h_0}}\{\max\{Z_i(r+1-d+h_0), Y_i(r)\}\}\\
 \end{eqnarray*}
}\normalsize

Since each node at hop distance $(h_0+1)$ transmits successfully at
least once in round $r,$ the transmission of each such node is decoded
successfully by some node at hop distance $h_0.$ Hence,
\begin{eqnarray*}
  \max_{i\in G_{h_0}}Y_i(r) & = & \max_{j\in
    G_{h_0+1}}T_j(r,t_{j,r}) \\
  & = & \max_{j\in \bigcup_{h_0+1\leq f\leq d}G_f} Z_j(r-d+h_0+1)
\end{eqnarray*}
where the second equality follows from the induction hypothesis.
Hence,
{\small
  \begin{eqnarray*}
    && \max_{i\in G_{h_0}}T_i(r+1,t_{i,r+1}) \ = \ \max\{
      \max_{i\in  G_{h_0}}Z_i(r+1-d+h_0), \\
    && \hspace{85pt} \max_{j\in \bigcup_{h_0+1\leq f\leq d}G_f}
      Z_j(r-d+h_0+1)\} \\ 
    && \hspace{80pt} =  \max_{j\in \bigcup_{h_0\leq f\leq d}G_f} Z_j(r-d+h_0+1)\} 
  \end{eqnarray*}
}\normalsize which proves the claim for hop distance $h_0$ for round
$(r+1).$ By induction, the claim is true for each $h$ and each round
$r>d-h.$ Therefore, the sink Node~$s$ correctly sets
$\mathcal{Z}(r-d)=\max\{Z_s(r-d),Y_s(r)\}.$ The delay of the protocol
is $d\tau=\Theta(\sqrt{n\log^3 n})$ slots.

As transmissions by different nodes are independent, the analysis in
the diffusion of phase I of \textsf{One-Shot MAX} carries over. The
probability that the computed value of $\mathcal{Z}(r)$ is incorrect
for any given round is upper bounded by $\frac{k}{n^\alpha}$ for any
constants $\alpha, k >0.$

\section{Discussion}
\label{sec:discussion}

The total number of transmissions (successful as well as unsuccessful)
in one execution of \textsf{One-Shot MAX} is
$\Theta(\frac{n^{3/2}}{\log^{3/2} n}).$ In \textsf{Pipelined MAX}, a
total of $\Theta(n\log n)$ transmissions are made per round. Note that
the corresponding number is $\Theta(n)$ with a coordinated protocol
for both cases. 

Our analysis can be extended to the case where the nodes use pure
Aloha as the MAC. We need to use a transmission rate rather than a
transmission probability. The success probabilities are calculated
similarly except that we now have a collision window that is twice the
packet length. All calculations are analogous.

It is fairly straightforward to show that in a noiseless,
structure-free broadcast network, the histogram can be computed in
$\Theta(n)$ slots w.h.p. In the noisy broadcast network, by a simple
modification of the protocol of \cite{Gallager88}, we can show that
the histogram can be computed in $\Theta(n\log \log n)$ slots w.h.p.

\bibliographystyle{plain}
\bibliography{isit08_arxiv}
\end{document}